\documentclass[letter]{jpsj3}
\usepackage{bm}

\newcommand{\rmd}{{\rm d}}
 
\title{
Majorana Fermions and  Z$_2$ Vortices on a Square Lattice
}

\author{
\name{
Kouta \surname{Ejima} and Takahiro \surname{Fukui}
} 
}
\inst{
\address{
Department of Physics, Ibaraki University, Mito 310-8512, Japan
} 
}
\abst{
We present a simple model of Majorana fermions on a square lattice, and study 
zero-energy states due to Z$_2$ vortices. 
We show the relationship between the Chern number of the ground state and 
the number of zero-energy states through numerical calculations for finite systems.
We also discuss the same relationship for the bulk system by observing the change in 
the spectral asymmetry as well as the lattice index.
}

\kword{Majorana fermions, Z$_2$ vortex, Chern number, overlap Dirac operator, 
Ginsparg-Wilson relation
}

\begin{document}
\maketitle

Quantum statistics of identical particles is one of the most basic properties 
in the study of many-body systems. 
Among some unusual statistics of elementary excitations
known so far,
Majorana fermions obeying non-Abelian statistics are of much current interest and are expected to play a crucial role in fault-tolerant quantum computations 
\cite{Kitaev03,Kitaev06,Nayak:2008zza}.
Since the field theory of Majorana fermions \cite{Ginsparg89} describes
the critical point of the Ising model, they are often called Ising fermions.
Kitaev proposed a solvable spin model on a honeycomb lattice, in which spins can be mapped to
Majorana fermions.\cite{Kitaev06}
A $p_x+ip_y$ superconductor is another well-known example of them \cite{Volovik99,ReaGre00}.
These models belong to the same universality class D for 
topological superconductors \cite{SchnyderRFL:08,Kitaev:08}, and conversely,
superconductors in class D may be regarded as Majorana fermion systems.
\cite{Sato03,CJNPS10}

Defined on a lattice, the Kitaev model provides not only various numerical results
\cite{Lahtinen11} but also 
intriguing notions such as the Majorana number, which relates
the existence of the zero-energy states in the Z$_2$ vortex sectors
to the Chern number of the ground state \cite{Kitaev06}. 
The purpose of this paper is to present a simpler minimal lattice model written directly 
by Majorana fermions and defined on a conventional square lattice. 
%
This model is intimately related to the Wilson-Dirac operator in the lattice gauge theories.
The development in this field concerning a chiral anomaly on a lattice
 \cite{Neuberger98,Luescher98}
yields a notion of the spectral flow and resultant  
Z$_2$ index theorem for the zero-energy state of the Wilson-Dirac Hamiltonian.

Hamiltonian density is defined by
\begin{alignat}1
{\cal H}=-i\frac{1}{2}\gamma_\mu(\nabla_\mu+\nabla_\mu^*)
+\gamma_3\left(m+\frac{b}{2}\nabla_\mu^*\nabla_\mu\right) ,
\label{HamDen}
\end{alignat}
where $\mu=1,2$
and the lattice derivatives are defined by
\begin{alignat}1
a\nabla_\mu \psi(x)&=U_\mu(x)\psi(x+a\hat\mu)-\psi(x) ,
\nonumber\\
a\nabla_\mu^* \psi(x)&=\psi(x)-U_\mu^*(x-a\hat\mu)\psi(x-a\hat\mu) 
\label{DifOpe}
\end{alignat}
with the lattice constant $a$ and the unit vector $\hat\mu$ in the $\mu$
direction.
The link variable $U_\mu(x)$ is defined on the link between
$x$ and $x+a\hat\mu$. 
Let us choose the $\gamma$-matrices as
$
\gamma_1=\sigma^x,
\gamma_2=\sigma^z,
$ and $\gamma_3=-\sigma^y
$,
and restrict the link variables as $U_\mu(x)=\pm1$. Then, it turns out that
\begin{alignat}1
{\cal H}^*=-{\cal H},
\label{SymHam}
\end{alignat}
which means that the Hamiltonian is purely imaginary and antisymmetric.
This enables us to define the following Majorana fermion model on the square lattice:
\begin{alignat}1
H=&a^2\sum_x \psi(x){\cal H}\psi(x)
\nonumber\\
=&\sum_j\frac{-i}{2}
\left(U_{\mu,j}c_j\gamma_\mu c_{j+\hat\mu}-U_{\mu,j}^*c_{j+\hat\mu}\gamma_\mu c_j\right)
+M\sum_j c_j\gamma_3c_j
\nonumber\\
&+\frac{B}{2}\sum_j\left(U_{\mu,j}c_j\gamma_3c_{j+\hat\mu}+U_{\mu,j}^*c_{j+\hat\mu}\gamma_3c_j
-2c_j\gamma_3c_j\right) ,
\label{Ham}
\end{alignat}
where $\psi(x)$ in the first line denotes a Majorana fermion operator on the site $x$ obeying 
$\{\psi(x),\psi(y)\}=\delta_{xy}/a$. In the next lines, we have 
introduced the dimensionless operator $c_j=a^{1/2}\psi(x)$ as well as dimensionless parameters
$M=ma$ and $B=b/a$, where $j=x/a$ is the set of integers indicating the lattice point. 


It may be noted that 
the Hamiltonian density (\ref{HamDen}) is related to,  as will be discussed momentarily, 
the Wilson-Dirac operator on the square lattice,
$D_W=\left[\gamma_\mu(\nabla_\mu+\nabla_\mu^*)-b\nabla^*_\mu\nabla_\mu\right]/2$,
where the Laplacian term is the famous 
Wilson term introduced for the purpose of avoiding species doubling.
Remarkably, the Hamiltonian (\ref{Ham})  is intimately related to condensed matter physics,
which has been used as a prototype of topological insulators.
\cite{BHZ:06,RHV:10,QHZ10,Qi:11,FHNQ:11} 
Although all these studies focus on the Dirac fermions on a lattice,
the Hamiltonian density (\ref{HamDen}) serves as a simple Majorana fermion model, similarly to the Kitaev model. 

Let us begin with discussions on the Chern number of the ground state without vortices.
To this end, we set $U_\mu(x)=1$ for all $x$.
The Fourier transformation yields
\begin{alignat}1
{\cal H}_k&= 
\sum_\mu s_\mu\gamma_\mu+\left[M+B\sum_\mu(c_\mu-1)\right]\gamma_3
\equiv X_A\gamma_A,
\label{HamMom}
\end{alignat}
where $s_\mu$ and $c_\mu$ ($\mu=1,2$) are lattice momenta defined 
by $s_\mu\equiv\sin k_\mu$ and $c_\mu\equiv\cos k_\mu$, and $A$ runs $A=1,2,3$.
The energy spectrum is given by $\pm R$, which follows from ${\cal H}_k^2=R^2\bm1$
with $R=\sqrt{\sum_AX_A^2}$.
The first Chern number $c$ characterizing the negative energy state
of eq. (\ref{HamMom}) may be computed using  
the projection operator $P_k\equiv (1-{\cal H}_k/R)/2=(1-\hat X_A\gamma_A)/2$ such that
\begin{alignat}1
c&=\frac{i}{2\pi}\int{\rm tr}\,P(\rmd P)^2
=\frac{1}{4\pi}\int \Omega ,
\label{CheNum}
\end{alignat}
where $\hat X_A$ is defined by $\hat X_A\equiv X_A/R$, which is regarded as the coordinates 
of ${\rm S}^2$ because of  $\sum_A\hat X_A^2=1$.The exterior derivative $\rmd$ is with respect to the momentum 
defined by $\rmd=\rmd k_\mu\partial_{k_\mu}$, and $\Omega$ is a two-form given by 
\begin{alignat}1
\Omega&\equiv\frac{1}{2!}\epsilon_{ABC}\hat X_A\rmd\hat X_B\rmd\hat X_C
\nonumber\\
&=\frac{c_1c_2}{R^3}
\left[M+B\sum_\mu(c_\mu-1)+B\sum_\mu s^2_\mu /c_\mu\right] \rmd^2k .
\label{Ome}
\end{alignat}
As discussed by Fujiwara {\it et al.} \cite{FNS02}, $k_\mu\rightarrow \hat X_A$ defines the 
mapping  $f:{\rm T}^2\rightarrow {\rm S}^2$, and the integral of $\Omega$
over the Brillouin zone in eq. (\ref{CheNum}) is simply the degree of the mapping $f$
times the volume of $S^2$, i.e., $4\pi$.
To calculate the degree of the mapping $f$,
let us consider the point $\hat{\bm X}=(0,0,1)$ on ${\rm S}^2$. The preimage of this point under $f$ 
is the set of points $k_\mu=0,~\pi$. 
Note that 
$\hat X_3={\rm sgn}(M)$ for $\bm k=(0,0)$, ${\rm sgn}(M-2B)$ for $(0,\pi)$ and $(\pi,0)$,
and ${\rm sgn}(M-4B)$ for $(\pi,\pi)$.
Taking the sign of the prefactor $c_1c_2$ in eq. (\ref{Ome}) into account, we finally conclude that 
the Chern number is $c={\rm sgn}(B)$ for $0<M/(2B)<1$, $c=-{\rm sgn}(B)$ for $1<M/(2B)<2$, 
and $c=0$ otherwise.

\begin{figure}[h]
\begin{center}
\begin{tabular}{cc}
\includegraphics[width=0.35\linewidth]{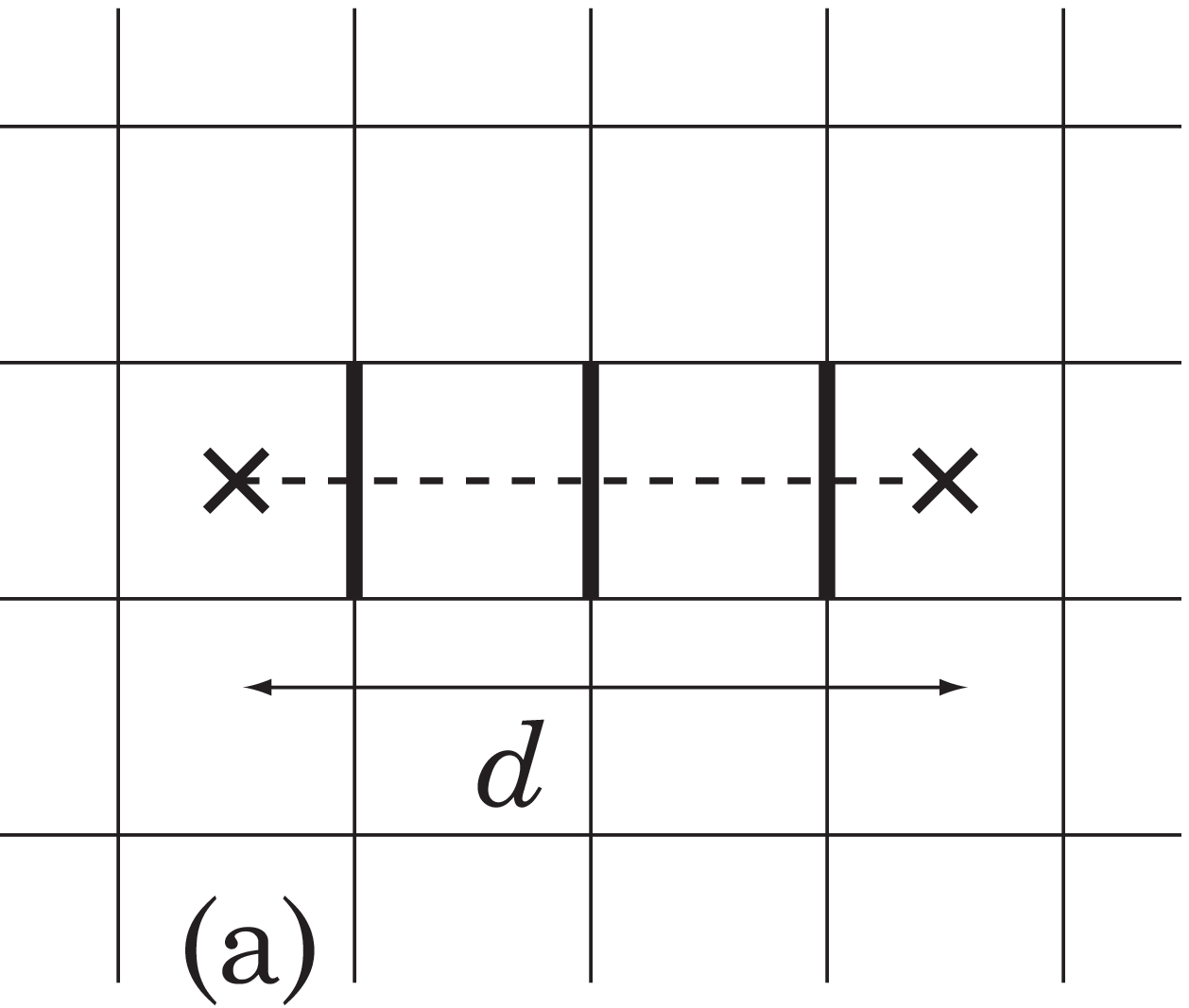}&
\includegraphics[width=0.35\linewidth]{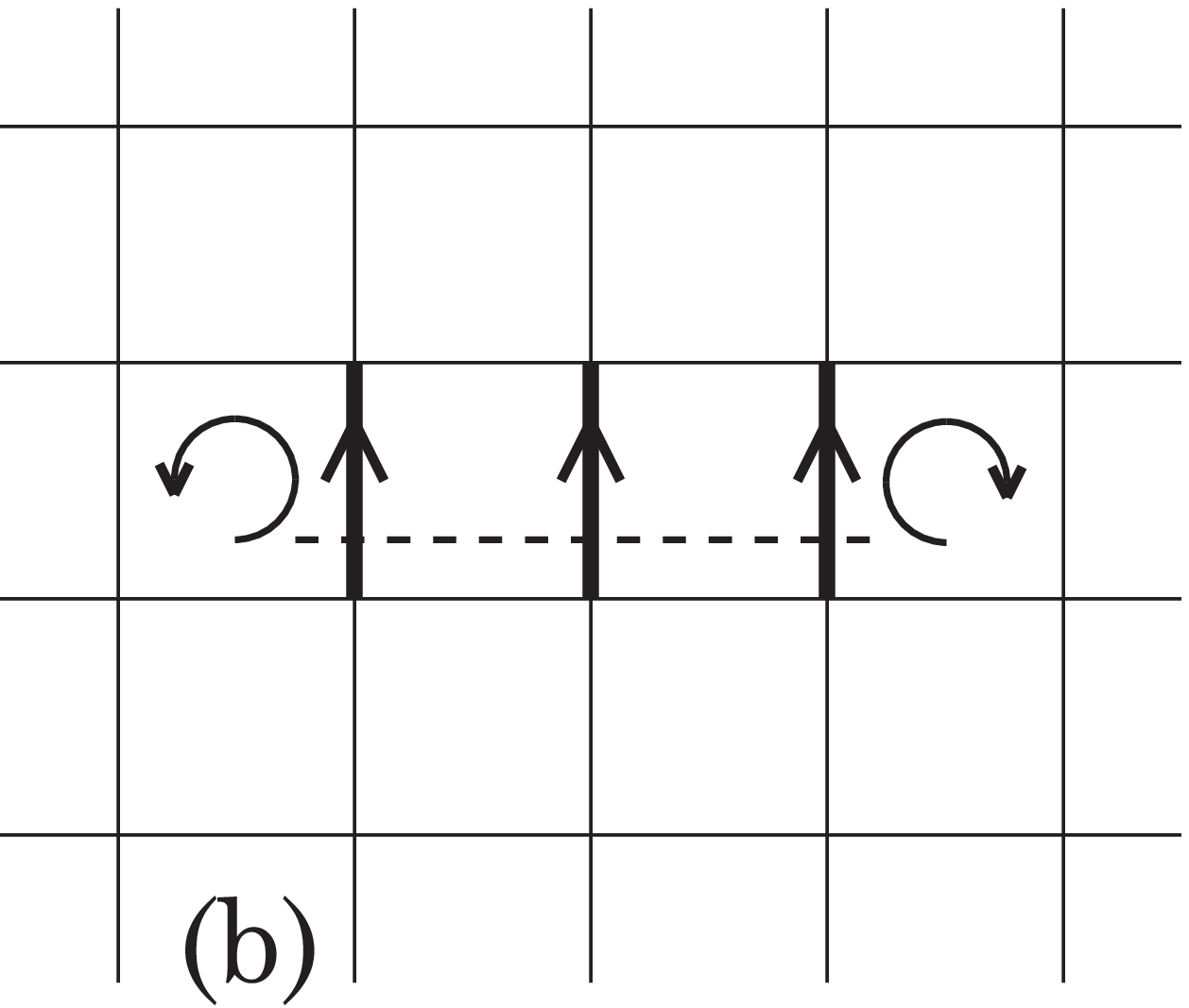}
\end{tabular}
\caption{
(a) Two Z$_2$ vortices separated by distance $d$. On the thick links crossing the dashed line,
$U_\mu=-1$ link variables are set.
The crosses represent the positions of Z$_2$ vortexes.
(b) Z$_2$ link variable with $U_\mu=-1$ in (a) is replaced by a  generic U(1) variable $e^{i\phi}$.
Because of the complex link variables, an arrow
indicating the positive $\phi$ direction is attached to each link, 
and the curved arrows show the positions of U(1) vortexes.
}
\label{f:VorCon}
\end{center}
\end{figure}

We now begin the discussions on the zero-energy states around $Z_2$ vortices.
For numerical calculations of the finite size systems, we set two Z$_2$ vortices separated by
distance $d$, as illustrated in Fig. \ref{f:VorCon}(a). 
\begin{figure}[h]
\begin{center}
\begin{tabular}{cc}
\includegraphics[width=0.47\linewidth]{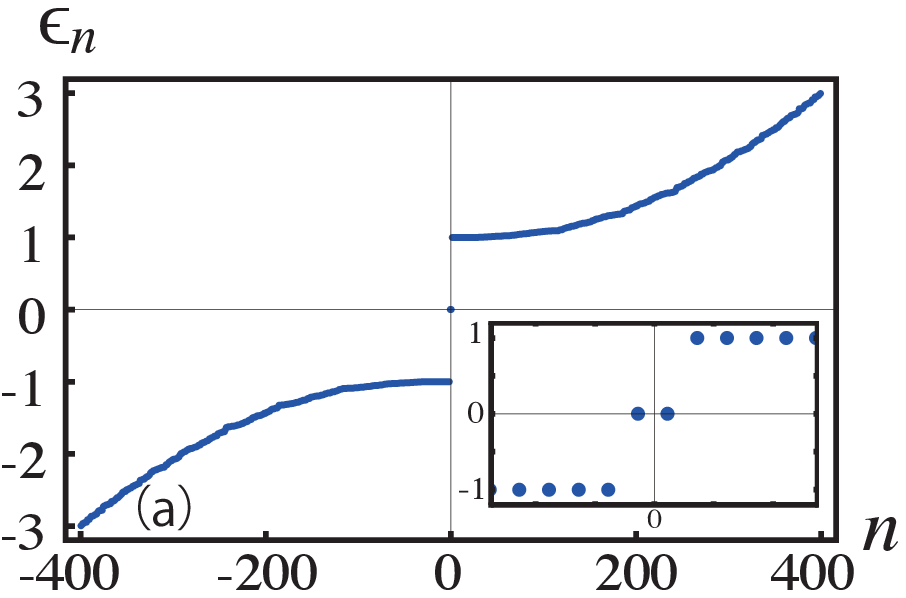}&
\includegraphics[width=0.47\linewidth]{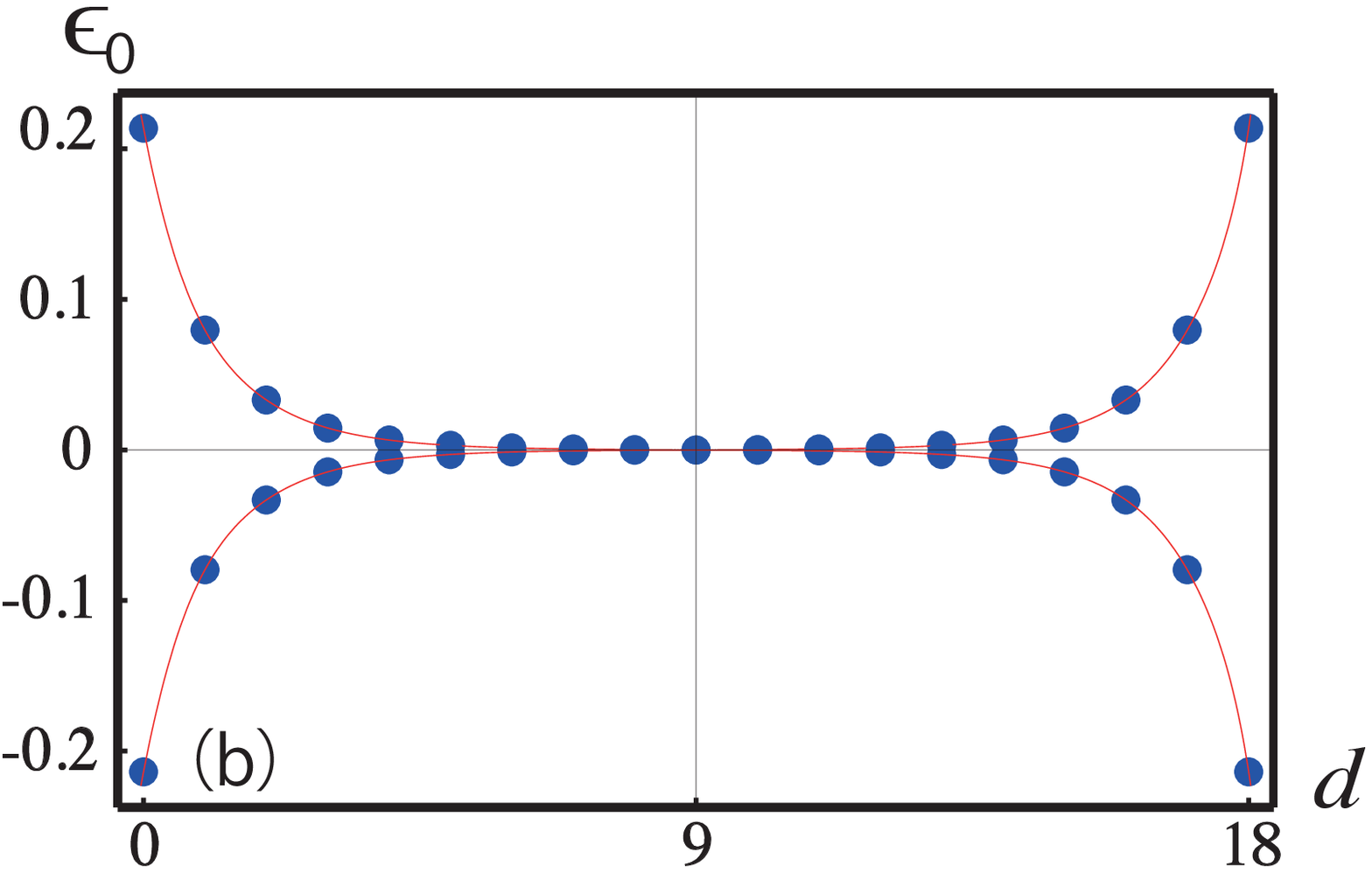}
\vspace*{-2mm}\\

\includegraphics[width=0.47\linewidth]{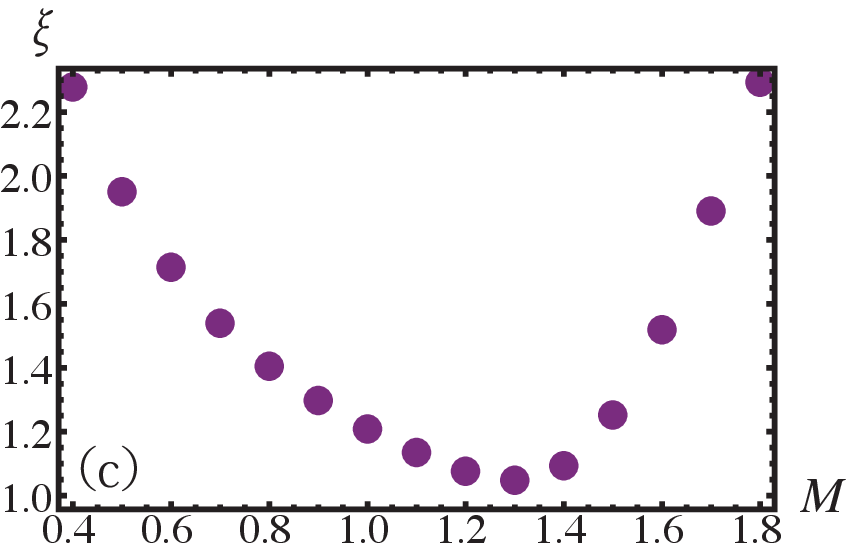}&
\includegraphics[width=0.46\linewidth]{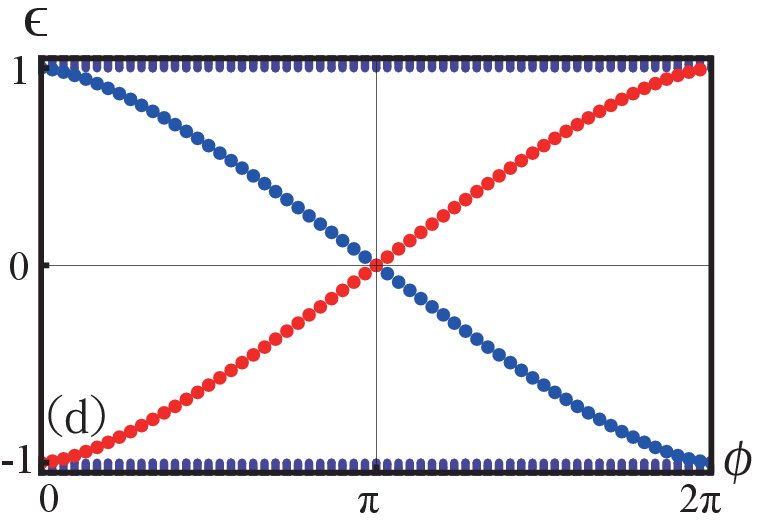}
\vspace*{-3mm}\\

\end{tabular}
\caption{(Color online)  
(a) Spectrum of finite size system ($20\times20$) under periodic boundary conditions.
The distance between vortices is $d=9$, and other 
parameters used are $M=1$ and $B=1$ ($c=1$ for ground state).
Inset: Spectrum around the zero energy.
(b) Energy $\pm\epsilon_0$ as a function of $d$.
(c) Coherence length $\xi$ as a function of $M$.
(d) Spectral flow  as a function of $\phi$ 
when all Z$_2$ link variables with $U_\mu=-1$ are replaced by $U_\mu=e^{i\phi}$.
}
\label{f:Spe}
\end{center}
\end{figure}
Since
the spectrum is symmetric with respect to the zero-energy owing to property (\ref{SymHam}),
we denote the eigenenergies as $\pm \epsilon_n$ $(n=0,1,2,\cdots)$.
Figure \ref{f:Spe}(a) shows the spectrum of the system that has a $c=1$ ground state.
There appear two approximately zero-energy states in the presence of well-separated Z$_2$ vortices.
However, for smaller $d$, these states have finite energies, $\pm\epsilon_0$, owing to 
their interactions. We show in Fig. \ref{f:Spe}(b) the energy $\pm\epsilon_0$ as a 
function of the distance $d$. 
The energy $\epsilon_0$ can be well parameterized as \cite{Lahtinen11}
\begin{alignat}1
\epsilon_0=\Delta e^{-d/\xi},
\label{CorLen}
\end{alignat} 
where $\Delta$ is of the order of the bulk energy gap and $\xi$ is the coherence length
between two Z$_2$ vortices. This behavior is in sharp contrast to the Kitaev model, 
which shows an oscillation of $\Delta$ as a function of $d$.\cite{Lahtinen11}
Since the coherence length generically depends on the model parameters, we plot 
$\xi$ as a function of $M$ in Fig. \ref{f:Spe}(c).

We next show the zero-energy states from the point of view of 
spectral flow, by extending the Z$_2$ link variable
with $U_\mu=-1$ to the generic U(1) variable $U_\mu=e^{i\phi}$.
As illustrated in Fig. \ref{f:VorCon}(b),
this yields a pair of a U(1) vortex and an antivortex, and exactly at $\phi=\pi$, they become
the two Z$_2$ vortices discussed above. We show in Fig. \ref{f:Spe}(d) the spectral flow as a function
of $\phi$. 
Starting at $\phi=0$, which corresponds to the ground state without vortices, and
increasing $\phi$ adiabatically, two eigenvalues
approach zero, and at $\phi=\pi$, they form exponentially degenerate zero-energy states.
The increasing (decreasing) state  
denoted by red (blue) points is associated with the vortex (antivortex).
This behavior suggests that if there is a single vortex in the infinite bulk system,
only one state (in the present case, only the increasing state denoted by red points)
crosses the zero energy without a level repulsion, and hence, 
we have just one zero-energy state sitting exactly on the zero energy at $\phi=\pi$.

We have so far shown some results of the model with a $c=1$ ground state.
We mention in passing that contrary to this case, 
we numerically confirm that the model with a
$c=0$ ground state shows no zero-energy states even in the presence of distant vortices.

We now elucidate a true zero-energy state associated with a single Z$_2$ vortex in the bulk system. 
We here remind ourselves that the Hamiltonian density eq. (\ref{HamDen})
is intimately related to the Wilson-Dirac operator. 
From the point of view of lattice fermions, 
this operator is free from the species doubling owing to the Wilson term, 
but at the cost of a loss of chiral symmetry.
Recent developments, however,  enable us to define a variant of chiral symmetry on a lattice.
A key property of such developments is the Ginsparg-Wilson (GW) relation, which has been recently
realized as the overlap Dirac operator.

The GW relation is given, in the case of two dimensions, by 
$\gamma_3D+D\gamma_3=aD\gamma_3 D$, where $D$ is a generic Dirac operator obtained by
the renormalization group for a lattice fermion \cite{GinWil:82}. 
In this equation, 
the r.h.s  means broken chiral symmetry since the lattice constant is finite.
Suppose that $D$ satisfies the GW relation and that $\hat\gamma_3$ is a variant of 
$\gamma_3$ on a lattice defined by $\hat\gamma_3=\gamma_3(1-aD)$.
Then, one can show that $\gamma_3D+D\hat\gamma_3=0$ owing to the GW relation.
This can be regarded as chiral symmetry on a lattice. 
A concrete operator satisfying the GW relation was constructed and called the
overlap Dirac operator.\cite{Neuberger:98}
It is defined by the use of the Wilson-Dirac operator as follows.
Let $A$ be an operator defined by $A=M-aD_W$. Then, the overlap operator $D$ is defined by
$D=\left[1-A(A^\dagger A)^{-1/2}\right]/a$.
With the use of $D$, the chiral anomaly can be reproduced on a lattice
such that $q(x)={\rm tr}\,\left[\gamma_3\left(1-aD(x,x)/2\right)\right]/a^2$, 
which also leads to the index theorem on a lattice. \cite{HLN:98}

With appropriate choices of the $\gamma$-matrices for ${\cal H}$ in eq. (\ref{HamDen}) 
and $D_W$ as well, we have $a{\cal H}=\gamma_3A$, and hence the overlap Dirac operator
can be expressed as 
\begin{alignat}1
D=\frac{1}{a}\left(1-\gamma_3{\cal H}/\sqrt{{\cal H}^2}\right) .
\label{D}
\end{alignat}
The chiral anomaly can then be expressed by 
$q(x)={\rm tr}\,\left(\gamma_3+{\cal H}/\sqrt{{\cal H}^2}\right)/(2a^2)$.
This has a topological meaning through the index theorem mentioned above.
This motivates us to study
\begin{alignat}1
Q&\equiv a^2\sum_{x} q(x)
=\frac{1}{2}{\rm Tr}\,\frac{{\cal H}}{\sqrt{{\cal H}^2}},
\label{DefQ}
\end{alignat}
where ${\cal H}$ is the Hamiltonian density defined by eq. (\ref{HamDen}), and
in the last equality the $\gamma_3$-term independent of ${\cal H}$ has been omitted,
since it vanishes.
Alternatively, this quantity can be viewed as a spectral asymmetry,
$Q={\rm Tr}\,(P_+-P_-)/2$, where
$P_\pm=(1\pm{\cal H}/\sqrt{{\cal H}^2})/2$ represents the projection operator
to the positive or negative energy state.

To explore the zero-energy state around a vortex, we set a single Z$_2$ vortex at the origin 
accompanied by an infinite sequence of $-1$ links from the origin to infinity.
In this case, eq. (\ref{DefQ}) is ill-defined, since ${\cal H}$ is expected to have 
a zero-energy state.
To overcome the difficulty, we utilize spectral flow again. Namely,  
we extend the link variables with $U_\mu(x)=-1$ to a generic U(1) variable,
$U_\mu(x)=e^{i\phi}$, to see spectral flow as a function of $\phi$.
This idea was originated by Roy \cite{Roy10}, who studied the zero modes
on the basis of the arguments of Laughlin \cite{Laughlin81} and Halperin.\cite{Halperin82}
In what follows, we will give an alternative proof of spectral flow, by calculating it
directly.

The extended Hamiltonian density, denoted by ${\cal H}(\phi)$,
becomes non-antisymmetric, and property (\ref{SymHam})  is modified as
\begin{alignat}1
{\cal H}^*(\phi)=-{\cal H}(-\phi) .
\label{GenSym}
\end{alignat}
For special cases $\phi=0,\pi$, and $2\pi$, the extended Hamiltonian describes the Majorana
fermions: ${\cal H}(0)$ and ${\cal H}(2\pi)$ are the original vortex-free Hamiltonian,
and ${\cal H}(\pi)$ is the Hamiltonian in the presence of a $Z_2$ vortex.
Of course, ${\cal H}(0)$ and ${\cal H}(2\pi)$ are identical. However, if the spectrum is followed 
as a function of $\phi$, spectral flow occurs and the states are generically rearranged. 
Even when ${\cal H}$ has zero eigenvalues at some $\phi$, in particular, at $\phi=\pi$, 
we can avoid them by adding some infinitesimal perturbations. For example,
around $\phi=\pi$, we extend $\phi$ into $\phi\rightarrow\phi+i\epsilon$, and then, 
the Hamiltonian density becomes non-hermitian and spectral flow bypasses the zero eigenvalues.
Thus, we can compute the spectral asymmetry (\ref{DefQ}) continuously from $\phi=0$ to $\phi=2\pi-0$.


Of course, it is difficult to calculate it 
directly for Hamiltonian (\ref{HamDen}). Remarkably,  
its general form  was determined by L\"uscher \cite{Luescher98} 
using only the cohomological arguments on the basis of
the locality, the gauge invariance, and the topological nature of the anomaly.
Some unknown parameters in this generic form were determined by direct calculation 
up to $a^0$, assuming a smoothly interpolated  gauge potential on the links. 
In the plane wave basis, \cite{FNS02} $q(x)$ of O$(a^0)$ can be computed from
\begin{alignat}1
q(x)
&=\frac{1}{2}\int\frac{d^2k}{(2\pi)^2}{\rm tr}\,
\frac{{\cal H}_k+\delta{\cal H}}{\sqrt{R^2+\{{\cal H}_k,\delta{\cal H}\}+\delta{\cal H}^2}} ,
\end{alignat}
where 
the leading term ${\cal H}_k$ is defined by eq. (\ref{HamMom}), 
and $\delta{\cal H}$ is the remaining difference operators defined by the relation
$e^{-ikx}{\cal H}e^{ikx}={\cal H}_k+\delta{\cal H}$.
Note that ${\rm tr}\,\gamma_A\gamma_B\gamma_C=2i\epsilon_{ABC}$, the leading contribution in eq. (12)
is given by
\begin{alignat}1
q(x)&=-\frac{1}{4}\int\frac{d^2k}{(2\pi)^2}\frac{1}{R^3}{\rm tr}\,
{\cal H}_k\delta{\cal H}^2 
\nonumber\\
&=\frac{c}{2\pi}F_{12}+O(a^1),
\label{SpeAsy}
\end{alignat}
where $c$ is the Chern number in eq. (\ref{CheNum}), and 
$F_{12}$ is the field strength associated with the U(1) gauge potential. 
It should be stressed here that the above $F_{12}$ is the field strength 
of order $a^0$, i.e., in the continuum limit, assuming 
that the U(1) link variable can be parameterized by a smooth gauge potential.
Having a known leading order, we obtain the exact $q(x)$, 
according to the L\"uscher's arguments, as 
\begin{alignat}1
q(x)=\frac{c}{2\pi}F_{12}(x)+\partial_\mu^* k_\mu(x) , 
\label{SpeAsyFul}
\end{alignat} 
where $k_\mu(x)$ is a gauge-invariant local current, $\partial_\mu^*$ is 
$\nabla^*$ with $U_\mu=1$ in eq. (\ref{DifOpe}), and $F_{12}$ is the field strength on a lattice: 
\begin{alignat}1
a^2F_{12}(x)=-i\ln U_1(x)U_2(x+a\hat1)U_1^*(x+a\hat2)U_2^*(x) .
\label{FieStr}
\end{alignat}  
It should be noted that 
the current $k_\mu$ is a lattice artifact,\cite{FNS02} and vanishes in the continuum limit $a\rightarrow0$.

Now, let us explore the zero mode in the presence of a single Z$_2$ vortex.
To this end, we investigate the spectral flow as a function of $\phi$.
First, when $\phi=0$, the model includes no vortices.
We readily see $F_{12}=0$. 
Note also 
the fact that the Hamiltonian has a well-defined continuum limit, implying
that $a^2\sum_x\partial_\mu k_\mu^*=0$. Therefore, we conclude that 
$Q=0$ when $\phi=0$. 
Let us now increase $\phi$ adiabatically from $\phi=0$. 
We specify the branch of the logarithm
in eq. (\ref{FieStr}) such that $0\le a^2F_{12}<2\pi$.
Then, the field strength (\ref{FieStr}) becomes 
$a^2F_{12}=\phi$ at the plaquette of the vortex and $a^2F_{12}=0$ elsewhere.
Unfortunately, for a finite $\phi$, it is impossible to take the continuum limit of the model
for which
it is difficult to calculate $k_\mu$ and therefore, $Q$.
Finally, when $\phi\rightarrow2\pi-0$, it follows from eq. (\ref{SpeAsyFul}) that
$Q\rightarrow c$, since in this limit, the continuum limit is well defined and 
the contribution from the current $k_\mu$ should vanish again. 
Therefore, we obtain the spectral flow 
\begin{alignat}1
\delta Q\equiv Q(2\pi-0)-Q(0)=c .
\label{MaiThe}
\end{alignat}
More generically, if one considers the case of $n_\pm$ vortices with $\pm$ vorticity,
$\delta Q=c\delta n$, where $\delta n=n_+-n_-$.

For simplicity, we restrict our discussions to the case of $c=1$.
Then, $\delta Q=1$, which implies that one negative state at $\phi=0$ flows towards the 
positive energy as $\phi$ increases, and finally at $\phi=2\pi$, it becomes a positive energy state
above the gap.
Let $\epsilon(\phi)$ be the eigenenergy of such a state.
Then, the symmetry (\ref{GenSym}) guarantees that the spectral flow is symmetric with respect to
$\phi=\pi$, $\epsilon(\phi)=-\epsilon(2\pi-\phi)$, 
and therefore, $\epsilon(\pi)=0$.
This state is the zero-energy state due to a Z$_2$ vortex.
On the other hand, if the Chern number of the ground state is trivial, $c=0$, the spectral
flow does not occur and no zero-energy state appears. 
All these observations are consistent with the numerical calculations.
More generically, if $\delta Q=\mbox{even}$, an even number of states flow and they cross the 
zero energy at generic $\phi$ owing to the level repulsion at $\phi=\pi$, 
whereas if $\delta Q=\mbox{odd}$, generically, only one state should cross
the zero energy exactly at $\phi=\pi$. 
Therefore, the number of zero-energy states associated with Z$_2$ vortices is given by 
$\delta Q$ modulo 2.

Although we have investigated eq. (\ref{DefQ}) as the spectral asymmetry,
it should also have the meaning of the index of the Hamiltonian, and eq. (\ref{MaiThe})
can be interpreted as the topological change of the index at $\phi=\pi$.
To explore the index, we note that only the zero-energy states of the Dirac operator $D$ 
in eq. (\ref{D}) contribute to the index (\ref{DefQ}). \cite{FujSuz04}
In what follows, for simplicity, we restrict our discussions to the nontrivial case 
$0<M<2B$, giving $c=1$ for the ground state. 
Then, only one mode around $\bm k\sim(0,0)$ becomes massless in eq. (\ref{D}), 
whereas other doublers $\bm k\sim(0,\pi), (\pi,0), (\pi,\pi)$ are still massive.
Therefore, the massless mode around zero momentum is expected to contribute to the index. 
To compute the index, we must introduce the flux $\phi$ for this mode.
In the lattice model, the flux is located at one plaquette, which makes it difficult to
consider the continuum limit of the model. 
Instead, we consider a flux defined in a wider region on a two dimensional space, which should give 
the same index as long as the total flux is kept unchanged. 
Thus, we explore the effective continuum Hamiltonian that we expect gives the index (\ref{DefQ}),
\begin{alignat}1
{\cal H}_{\rm c}=-i\gamma_\mu(\partial_\mu -iA_\mu),
\label{DirConMod}
\end{alignat}
where we assume that the gauge potential is given by \cite{JackiwRossi81}
\begin{alignat}1
A_\mu(x)=\epsilon_{\mu\nu}\hat x_\nu A(r) 
\end{alignat}
with the asymptotic form
\begin{alignat}1
A(r)\rightarrow 
\left\{
\begin{array}{ll}
0 \qquad & (r\rightarrow0)\\
-\displaystyle{\frac{\phi}{2\pi r}} & (r\rightarrow\infty)
\end{array}
\right. .
\end{alignat}
Integrated over 2D, this gauge potential indeed yields the total flux $\phi$. 

Suitable unitary transformation enables us to choose $\gamma_1=\sigma^x$, $\gamma_2=\sigma^y$, 
and $\gamma_3=\sigma^z$. 
In this representation, let us obtain the zero-energy states ${\cal H}_{\rm c}\psi=0$, which is explicitly
given by each chiral component $\pm$ such that 
\begin{alignat}1
\left(\partial_r\pm\frac{i}{r}\partial_\theta \mp A(r)\right)\psi_\pm(r,\theta)=0 ,
\end{alignat}
where $\psi_\pm(r,\theta)$ is the zero-energy wavefunction with chirality $\pm$,
i.e., $\gamma_3\psi_\pm=\pm\psi_\pm$.
Set $\psi_\pm(r,\theta)=e^{\pm im\theta}\psi_\pm (r)$. Then, the equation is reduced to 
\begin{alignat}1
\left(\partial_r-\frac{m}{r} \mp A(r)\right)\psi_\pm(r)=0 .
\end{alignat}
We obtain the solutions of this equation with the following 
power law behavior at the boundaries,
\begin{alignat}1
\psi_\pm\sim
\left\{
\begin{array}{ll}
r^m\quad & (r\rightarrow0)\\
r^{m\mp\phi/(2\pi)} & (r\rightarrow\infty)
\end{array}
\right. .
\end{alignat}
The normalizability of the wavefunction requires
\begin{alignat}1
-1<m<\pm\phi/(2\pi)-1.
\end{alignat}
When $\phi=0$, there are no solutions, but when $\phi$ exceeds $\pi$, there appears the solution
$m=-1/2$ in the chirality $+$ sector. 
This wavefunction is antiperiodic (double-valued),\cite{JackiwRossi81}
$\psi_+(\theta= 2\pi)=-\psi_+(\theta=0)$.
The change in its index from 0 to 1 (\ref{MaiThe}) owing to spectral flow 
may thus be interpreted by this effective model. 
Although the model (\ref{DirConMod}) corresponds to the Dirac fermion with U(1) gauge potential, 
it can be written as Majorana doublet fermions with an O(2) gauge potential.

\begin{acknowledgments}
The authors would like to thank T. Fujiwara for fruitful discussions.
This work was supported in part by a Grant-in-Aid for Scientific Research (No. 21540378) from the Japan Society for the Promotion of Science (JSPS)
 and by 
the ``Topological Quantum Phenomena'' 
Grant-in Aid for Scientific Research 
on Innovative Areas (No. 23103502) from the Ministry of Education, Culture, Sports, Science and Technology of Japan (MEXT).


\end{acknowledgments}


\end{document}